\documentclass[conference]{IEEEtran}
\usepackage{graphicx}
\usepackage{amsmath}
\usepackage{amssymb}
\usepackage{multirow}
\usepackage{cite}
\usepackage[utf8]{inputenc}
        \usepackage[T1]{fontenc}
\usepackage{amsthm}
\usepackage{algorithm,algorithmicx}
\usepackage{algpseudocode}

\usepackage{hyperref}
\usepackage{cleveref}
\usepackage{epstopdf}
\usepackage{color}
\usepackage{algorithm}
\usepackage{algpseudocode}
\newtheorem*{proposition*}{Proposition}
\begin{document}

\newcommand{\mat}{\mathbf}
\renewcommand{\vec}[1]{\bar{\mat{#1}}}
%
\title{Online Time Sharing Policy in Energy Harvesting Cognitive Radio Network with Channel Uncertainty}

\author{\IEEEauthorblockN{Kalpant~Pathak, Prachi Bansal, and Adrish~Banerjee}
\IEEEauthorblockA{Department of Electrical Engineering, Indian Institute of Technology Kanpur, Uttar Pradesh, 208016\\
Email: \{kalpant, adrish\}@iitk.ac.in \vspace{-7mm}}

}

%

\maketitle

\begin{abstract}
This paper considers an energy harvesting underlay cognitive radio network operating in a slotted fashion. The secondary transmitter scavenges energy from environmental sources in half duplex fashion and stores it in finite capacity rechargeable battery. It splits each slot into two phases: \textit{harvesting phase} and \textit{transmission phase}. We model the energy availability at the secondary user as first order stationary Markov process. We propose a robust online transmission policy by jointly optimizing the time sharing between the two phases and transmit power of secondary user, which maximizes its average throughput by a given time deadline. We show the comparison of our proposed policy with the offline and myopic policies.

\end{abstract}

\IEEEpeerreviewmaketitle

\section{Introduction}
In recent years, there have been many developments for improving the bandwidth and energy efficiency of a wireless communication network. Energy harvesting cognitive radio network (EH-CRN) is one such solution which improves the bandwidth efficiency of the network while ensuring perpetual operation of the devices at the same time \cite{EH_survey_1,EH_survey_2,EH_CRN_survey}. In a CRN, a set of unlicensed users share the spectrum allocated to licensed users in a way such that the licensed user can achieve an acceptable quality of service (QoS). The unlicensed and licensed users are also known as secondary user (SU) and primary user (PU) respectively. Depending on the way of sharing, the CRN can be classified into three categories: interweave, overlay and underlay.

In underlay EH-CRN, the SUs and PUs coexist in an interference limited scenario and may harvest energy from the environmental sources. The secondary transmitter (ST) transmits its data using the spectrum allocated to PU while keeping acceptable interference at the primary receiver (PR). EH-CRN operating in underlay mode has been studied in \cite{underlay_pappas,underlay_duan,underlay_xu_multihop,underlay_xu,underlay_kalamkar,underlay_myopic,my_spcom}. We briefly summarize the related literature and present our main contribution.

In \cite{underlay_pappas} and \cite{underlay_duan}, authors considered an underlay EH-CRN with multipacket reception model. The SU transmits not only when PU is idle, it also transmits with a probability $p$ when PU is occupying the channel. Both works studied the stable throughput region and obtained optimal transmission probability $p$ maximizing the SU's throughput. However, \cite{underlay_pappas} considers an EH-PU whereas, \cite{underlay_duan} considers two different scenarios, EH-SU and, EH-PU and EH-SU. In \cite{underlay_xu_multihop}, authors considered an underlay EH-CRN with one PU and multiple EH-SUs. The SUs harvest RF energy from primary's transmission and use multihop transmission along with TDMA to transmit their own data. Authors jointly optimize the time and power allocation maximizing the end-to-end throughput. In \cite{underlay_xu}, authors considered a scenario where multiple EH-SUs communicate with an intended receiver using TDMA. The authors jointly optimize the power and time allocation which maximizes the sum rate of the SUs. In \cite{underlay_kalamkar}, a scenario is considered where a SU communicates with the receiver via multiple energy harvesting decode-and-forward relays while ensuring the outage probability of PU is below an acceptable threshold. The authors obtained the outage probability of the SU for Nakagami-$m$ channel in closed form. In \cite{underlay_myopic}, authors considered a single pair of PU and SU operating in underlay mode. The EH-SU operates in half duplex fashion and harvests energy from PU's transmission for the first fraction of the slot and then transmits its data in the remaining. Authors aim to obtain a myopic policy which optimizes the time sharing between the two phases and maximizes the sum rate of SU under PU's outage constraint. In \cite{my_spcom}, authors consider the same system model as in \cite{underlay_myopic} and maximize the sum throughput of the SU by jointly optimizing the time sharing and power allocation among the slots.

We consider an underlay EH-CRN where the PU has a reliable energy source and SU is equipped with a rechargeable battery and harvests energy from the environmental sources such as solar, vibration, RF etc. Similar to the system model presented in \cite{underlay_myopic} and \cite{my_spcom}, in our model, the SU operate in slotted half-duplex fashion, i.e., at any given time, SU can either harvest energy from the environment or transmit its data. We consider both the energy and channel uncertainties in our model which to the best of our knowledge, has not been studied in the literature in the context of EH-CRN. We model the uncertainty in energy harvesting process as a first order stationary Markov process as in \cite{learning_theoretic} and the estimated channel gains are assumed to have bounded uncertainty as in \cite{robust_xu,robust_zhang}. The main contributions of this paper are as follows:
\begin{itemize}
\item We propose a robust online time sharing policy taking energy arrival and channel uncertainties into consideration. We formulate the problem of maximizing secondary average sum throughput by a given time deadline (short-term throughput) subject to energy harvesting constraint of secondary transmitter (ST) and interference constraint of primary receiver (PR) as a finite horizon discrete-time Markov decision process (MDP) \cite{MDP_puterman}.
\item We solve the optimization problem using finite horizon stochastic dynamic programming (SDP) \cite{DP_bertsekas,MDP_puterman} and compared the performance of our proposed online policy with the myopic \cite{underlay_myopic} and  offline policy \cite{my_spcom}.
\item In addition, we also investigate the effects of various system parameters such as different channel conditions, radius of the uncertainty region, battery capacity and interference threshold at PR on the proposed time sharing policy.
\end{itemize}

The organization of the paper is as follows. System model is presented in section II which includes the energy arrival model, battery dynamics and channel uncertainty model. Problem formulation is presented in section III. We discuss the results in section IV and finally, we conclude the paper in section V.

\textit{Notations:} The bold faced symbol (e.g. $\mat{A}$) represents a matrix and with bar (e.g. $\bar{\mat{a}}$) represents a vector. $\bar{\mat{a}}\succeq\bar{\mat{0}}$ means that every element of vector $\bar{\mat{a}}$, $a_i$ is greater than or equal to 0.
\section{System Model}
This section presents our system model, which includes the description of underlay EH-CR system operating in slotted mode, energy arrival process, battery dynamics at the secondary transmitter (ST), and channel uncertainty model.
\vspace{-3mm}
\begin{figure}[!ht]
\centering
\includegraphics[width=0.7\linewidth]{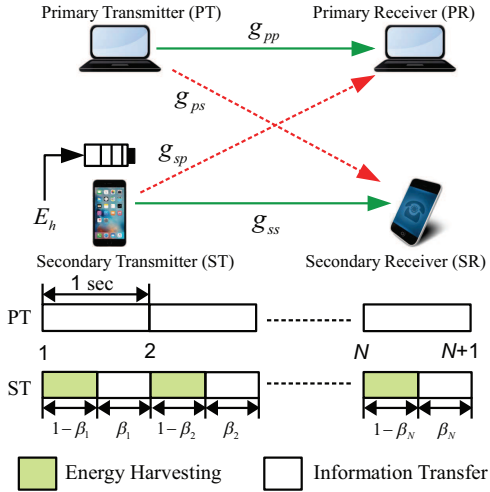}
\vspace{-1mm}
\caption{An underlay EH-CRN network}
\label{fig_system_model}
\vspace{-3mm}
\end{figure}
\subsection{Underlay EH-CRN Operating in Slotted Mode}
The underlay EH-CRN operating in slotted mode, is shown in Fig. \ref{fig_system_model}. In our model, the ST scavenges energy from the environment\footnote{As the power density of RF energy sources is too low \cite{RF_EH}, we do not consider RF energy harvesting in our work.} and stores it in rechargeable battery of finite capacity. In Fig. \ref{fig_system_model}, $g^i_{pp},g^i_{ps},g^i_{sp}$ and $g^i_{ss}$ represent the channel coefficients corresponding to PT-PR, PT-SR, ST-PR and ST-SR link in the $i$th slot respectively. Both PT and ST transmit simultaneously for $N$ slots, each of duration 1 second. In each slot, the PT uses a constant power $p_p$ for transmission. However, the secondary transmitter (ST) splits every $i$th slot into two phases: \textit{harvesting phase} and \textit{transmission phase} of duration $(1-\beta_i)$ and $\beta_i$ second respectively, where $0\leq\beta_i\leq1$. In \textit{harvesting phase} of $i$th slot, ST harvests energy and stores it in the battery of maximum capacity $B_{max}$ and then in the \textit{transmission phase}, it transmits its data to the secondary receiver (SR) with power $p_s^i$ Watt. The ST chooses its transmission power such that it causes at max $P_{th}$ Watt of interference at the PR.

We assume all the channel coefficients $g^i_{pp},g^i_{ps},g^i_{sp}$ and $g^i_{ss}$ to be i.i.d. zero mean complex Gaussian with variances $\sigma_{pp}^2,\sigma_{ps}^2,\sigma_{sp}^2$  and $\sigma_{ss}^2$ respectively. In $i$th slot, the instantaneous achievable throughput of the ST (in bps/Hz) is given as $
R_{i}\left ( \beta _{i},p_s^{i} \right )=\beta _{i}\log_{2}\left ( 1+\frac{\left | g_{ss}^{i} \right |^{2}p_s^{i}}{\sigma_n^2+\left | g_{ps}^{i} \right |^{2}p_{p} } \right ),\, \forall i,$
where $\left | g_{ss}^{i} \right |^{2}$ and $\left | g_{ps}^{i} \right |^{2}$ are the channel power gains of ST-SR and PT-SR link respectively, and $\sigma_n^2$ is the variance of zero mean additive white Gaussian noise (AWGN) at SR.
\subsection{Energy Uncertainty Model}
This section presents our model of energy uncertainty. We first present the energy harvesting process and then, we study the battery dynamics governed by the harvesting process.
\subsubsection{Energy Harvesting Process}
In our model, the ST has energy harvesting capability and harvests energy from the environmental sources. We assume that the ST operate in half duplex mode such that in the beginning of each slot, the SU first harvests energy from the environment with a rate $E_h^i$ J/s for some fraction of time, and then transmits its data in the remaining duration of the slot. 

In energy harvesting, the energy arrival time and amount are not known in advance and are random in nature. In order to capture this randomness, we model the process of energy as a first order stationary Markov process with $M_s$ number of states as in \cite{learning_theoretic}. The state transition probabilities are assumed to be known at the ST apriori. In practice, these transition probabilities can be estimated by observing the energy arrival pattern. At the beginning of $i$th slot, the ST harvests the energy from the environment at a harvesting rate $E_h^i$ which takes values from a finite set $\mathcal{E}=\{ e_1^h=0,e_2^h,\cdots,e_{M_s}^{h}\}$, where $e_1^h=0$ represents that no energy is harvested. In this paper, we consider $M_s=2$ without any loss of generality.
\begin{figure}[!h]
\centering
\includegraphics[width=0.65\linewidth]{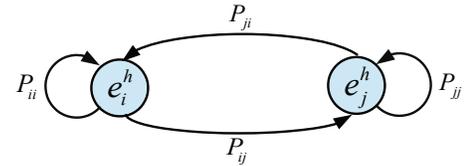}
\vspace{-1mm}
\caption{Two state Markov process}
\label{markov_process}
\vspace{-1mm}
\end{figure}

Fig. \ref{markov_process} shows a two state Markov process, where $P_{ij},\,i,j\in\{1,2\}$ are transition probabilities defined as
\begin{align*}
P_{ij}=\mathbb{P}(e_i^h\rightarrow e_j^h),\quad i,j\in\{1,2\}.
\end{align*}

We denote the state transition probability matrix by $\mathbf{T}$ such that $[\mat{T}]_{ij}=P_{ij}$, which is assumed to be known apriori. The state transition probability of the random variable $E_h^i$ is given as
\begin{align*}
\mathbb{P}\left (E_h^i\mid E_h^1,E_h^2,\cdots,E_h^{i-1} \right )=\mathbb{P}\left ( E_h^i\mid E_h^{i-1} \right ),\;\;  i=2,\ldots,N+1
\end{align*} 
\subsubsection{Battery Dynamics}
Since the energy harvesting process follows first order Markov process, so do the battery dynamics. The energy available in the battery at the beginning of each slot depends upon the energy harvested and energy consumed in the previous slot.

In the $i$th slot, the ST harvests energy for $1-\beta_i$ fraction of slot with a rate $E_h^i$ and then, transmits its data for $\beta_i$ fraction of the slot with power $p_s^i$. If $B_i$ is the state of the battery at the beginning of $i$th slot, we have
\begin{align}
0\leq \beta_ip_s^i\leq B_i,\quad\forall i. \label{eq:consumed_pwr_constr}
\end{align}

If $B_{max}$ denotes the capacity of the battery, the energy available in the battery at the beginning of $(i+1)$th slot can be expressed recursively as
\begin{align}
B_{i+1}=\textup{min}\left \{ B_i+(1-\beta _i)E_h^i-\beta _ip_s^{i}, B_{max} \right \}, \quad \forall i, \label{eq:battery_dynamics}
\end{align}
where $(1-\beta _i)E_h^i$ and $\beta _ip_s^{i}$ represent the harvested and consumed energies in the $i$th slot respectively. We assume $B_1=0$ without loss of generality. 

Both the harvested energy and battery state jointly determine the time sharing and transmit power in a slot. Therefore, we can form a new first order Markov process whose states are defined as the joint state of energy harvesting states and battery states. The $i$th state of this new Markov process, $Q_i$ can be defined as
\begin{align}
Q_i\triangleq \left\{
\begin{array}{ll}
B_1, & \text{for } i=1\\
(E_h^{i-1},B_i), & \text{for }i=2,\ldots,N\\
B_{N+1}, & \text{for }i=N+1
\end{array}
\right. 
\label{eq:new_markov_state}
\end{align}
The state transition probability of this new Markov process is given as
\begin{align*}
\mathbb{P}\left (Q_j\mid Q_1,\ldots,Q_{j-1} \right )=\mathbb{P}\left ( Q_j\mid Q_{j-1} \right ),\;\; j = 2,\ldots,N+1 
\end{align*}
\subsection{Channel Uncertainty Model}
We assume the coherence time of the fading channel to be equal to the slot length, i.e., the channel coefficients remain constant for each time slot but may vary from one slot to other. The ST can estimate the channel coefficients between itself to PR using channel reciprocity \cite{estimation_using_feedback}. However, due to practical constraints such as feedback delay or estimation errors, the estimated channel coefficients may be erroneous. Therefore we assume the CSI of PT-SR and ST-PR links to be imperfect with bounded uncertainty \cite{robust_xu,robust_zhang}. 

Under bounded uncertainty, the actual channel coefficients of PT-SR and ST-PR links can be written as
\begin{align*}
g_{ps}&=\hat{g}_{ps}+\Delta g_{ps}\\
g_{sp}&=\hat{g}_{sp}+\Delta g_{sp}
\end{align*}
where $\hat{g}$ and $\Delta g$ are the estimated channel coefficient and the estimation error respectively. Without assuming any statistical knowledge about the error, we bound the estimation error as $|\Delta g|\leq \varepsilon$, where $\varepsilon\geq0$ is the radius of the uncertainty region. We assume the estimated channel coefficients $\hat{g}_{ps}$ and $\hat{g}_{sp}$ to be zero mean complex Gaussian with variances $\hat{\sigma}^2_{ps}$ and $\hat{\sigma}^2_{sp}$ respectively.
\section{Problem Formulation}
\subsection{Online Policy}
The transmit power of ST, $p_s^i$ is controlled by the state of the new Markov process, $Q_i$ as well as the interference threshold at the PR, $P_{th}$. Our aim is to obtain optimal $\pmb{\beta}$ and $\bar{\mat{p}}_s$ which maximizes the worst case short-term average throughput of the ST considering the energy harvesting constraints of ST, interference threshold at PR, and imperfect CSI. The optimization problem is given as
\begingroup
\allowdisplaybreaks
\begin{subequations}
\begin{align}
\max_{\bar{\mat{p}}_s,\bar{\pmb{\beta}}}\min_{
\begin{array}{l}
|\Delta g_{ps}^i|\leq \varepsilon\\
|\Delta g_{sp}^i|\leq \varepsilon
\end{array}
}\quad & \mathbb{E}_{Q_2^N}\left\{\left [\sum_{i=1}^{N} R(\beta_i, p_s^i)  \right]| \mathbf{T} \right\} \label{eq:obj_orig}\\
\text{s.t.}\quad  0\leq \sum_{j=1}^{i}&\left( 1-\beta _{j} \right)E_{h}^{j}-\sum_{j=1}^{i}\beta_jp_s^{j}\leq B_{max},\;\;  \forall i \nonumber\\
&\hspace{-3mm} (\text{Energy causality constraint of ST})\label{eq:c1_orig}\\
&  0\leq \beta_ip_s^{i}\leq B_i,\;\;  \forall i \label{eq:c2_orig} \\
& \quad (\text{Consumed energy constraint of ST})\nonumber\\
& |\hat{g}^i_{sp}+\Delta g^i_{sp}|^2p_s^i\leq P_{th}, \quad \forall i \label{eq:c3_orig}\\
& \quad(\text{Interference constraint of PR})\nonumber\\
& \bar{\mat{0}}\preceq \bar{\pmb{\beta}}\preceq \bar{\mat{1}},\qquad \bar{\mat{p}}_s\succeq\vec{0} \label{eq:c4_orig}\\
& \quad(\text{Non-negativity constraint})\nonumber
\end{align}
\end{subequations}
\endgroup
where constraint (\ref{eq:c1_orig}) is the energy causality constraint. It states that in any slot, we can use as much energy as we have harvested upto that slot. The optimization problem in (\ref{eq:obj_orig})-(\ref{eq:c4_orig}) is a stochastic optimization problem where conditional expectation $\mathbb{E}_{Q_2^N}[\cdot|\mat{T}]$ is taken with respect to all possible values of state $Q_i,\,i=2,\ldots,N$ for a given state transition matrix $\mat{T}$. This optimization problem can be rewritten aiming for robust online policy as (See Appendix)
\begingroup
\allowdisplaybreaks
\begin{subequations}
\begin{align}
\max_{\bar{\mat{p}}_s,\bar{\pmb{\beta}}} \quad & \mathbb{E}_{Q_2^N}\left\{\left [\sum_{i=1}^{N} R_1(\beta_i, p_s^i)  \right]| \mathbf{T} \right\} \label{eq:obj_new}\\
\text{s.t.} \quad & \eqref{eq:c1_orig}, \, \eqref{eq:c2_orig},\, \eqref{eq:c4_orig} \label{eq:c1_new}\\
& (| \hat{g}_{sp}^{i}|^2 +2\varepsilon |\hat{g}_{sp}^{i}|+\varepsilon ^2)p_s^{i}\leq P_{th},\;\forall i  \label{eq:c2_new}
\end{align}
\end{subequations}
\endgroup
where $R_1(\beta_i, p_s^i)=\beta _{i}\log_{2}\left ( 1+\frac{\left|g_{ss}^{i} \right|^{2}p_s^{i}}{\sigma_n^2+(|\hat{g}_{ps}^{i}|^{2}+2\varepsilon |\hat{g}^i_{ps}|+\varepsilon^2)p_{p} } \right )$ is the worst case instantaneous achievable throughput of ST in $i$th slot.

The optimization problem (\ref{eq:obj_new})-(\ref{eq:c2_new}) can not be solved for each slot independently due to time coupled constraint (\ref{eq:c1_orig}). Therefore, we first rewrite the optimization problem in (\ref{eq:obj_orig})-(\ref{eq:c4_orig}) as a classical finite horizon MDP by reformulating the constraint (\ref{eq:c1_orig}) as (\ref{eq:battery_dynamics}) and combining the constraints (\ref{eq:c2_orig}) and (\ref{eq:c3_orig}). The optimization problem can now be rewritten as
\begingroup
\allowdisplaybreaks
\begin{subequations}
\begin{align}
\max_{\bar{\mat{p}}_s,\bar{\pmb{\beta}}}\quad & \mathbb{E}_{Q_2^N}\left\{\left [\sum_{i=1}^{N} R_1(\beta_i, p_s^i)  \right]| \mathbf{T} \right\} \label{eq:obj}\\
\text{s.t.}\quad & B_{i+1}=\textup{min}\left \{ B_i+(1-\beta _i)E_h^i-\beta _ip_s^{i}, B_{max} \right \}, \;\; \forall i \label{eq:c1}\\
& 0\leq p_s^i\leq \text{min}\left \{ \frac{B_i}{\beta_i},\frac{P_{th}}{|\hat{g}_{sp}^{i}|^{2}+2\varepsilon |\hat{g}^i_{sp}|+\varepsilon^2} \right \},\;\;\forall i \label{eq:c2}\\
& \bar{\mat{0}}\preceq \bar{\pmb{\beta}}\preceq \bar{\mat{1}},\qquad \bar{\mat{p}}_s\succeq\vec{0} \label{eq:c3}
\end{align}
\end{subequations}
\endgroup

The resulting optimization problem (\ref{eq:obj})-(\ref{eq:c3}) can now be solved optimally using finite horizon SDP \cite{MDP_puterman},\cite{DP_bertsekas}. The optimal values of optimization variables $\bar{\mat{p}}_s$ and $\bar{\pmb{\beta}}$ are obtained using backward induction method \cite{MDP_puterman} and are calculated in time reversal order. The SDP algorithm is given in Algorithm \ref{algo:SDP}.
\begin{proposition*}
The optimal last state of the newly formed Markov process is $Q^*_{N+1}=B_{N+1}=0$.
\end{proposition*}

The proposition states that by the end of the transmission, all the energy harvested would be consumed and in the end of last time slot, energy causality constraint will be satisfied with equality, i.e.,
\begin{align*}
\sum_{j=1}^{N+1}(1-\beta_j)E_h^j=\sum_{j=1}^{N+1}\beta_jp_s^j.
\end{align*}

It follows from the fact that it is always suboptimal to have some energy left in the battery at the end of the transmission.
\begin{algorithm}
\caption{SDP algorithm}
\label{algo:SDP}
\begin{algorithmic}
\State \textbf{Initialization:} Initialize $\mat{T}$, $Q_{N+1}=B_{N+1}=0$.
\State Set  $n\leftarrow N$
\State \textbf{Look up:}
\While{$n\neq1$}
\State Calculate $\mathbb{E}_{Q_n}\left\{\left [ R_1(\beta_n, p_s^n)  \right]| \mathbf{T} \right\}$ for all possible values of $Q_n,\, n=2,\ldots,N$. 
\State $n\leftarrow n-1$
\EndWhile
\State \textbf{Optimal $\bar{\mat{p}}_s$ and $\bar{\pmb{\beta}}$ using backward induction:}
\State set $n\leftarrow1$
\While{$n\neq N$}
\State given $Q_n=\{E_h^{n-1},B_n\}$, obtain
\State $[p_s^n,\beta_n]=\underset{{p_s^n,\beta_n}}{\arg\max}\,\mathbb{E}_{Q_n}\left\{\left [R_1(\beta_n, p_s^n)  \right]| \mathbf{T} \right\}$ from Look up.
\State $n\leftarrow n+1$
\EndWhile\\
\Return $\bar{\mat{p}}_s$ and $\bar{\pmb{\beta}}$
\end{algorithmic}
\end{algorithm}
\vspace{-3mm}
\subsection{Myopic Policy}
In the myopic policy, the SU aims to maximize its immediate throughput in each slot and therefore, it consumes all the harvested energy for transmission in the same slot. In this case, the throughput in each slot can be maximized by optimizing the time sharing parameter $\vec{\pmb{\beta}}$ only. Under myopic policy, the transmit power of SU in $i$th slot is given by $p_s^i=\frac{(1-\beta_i)}{\beta_i}E_h^i$. The optimization problem for robust myopic policy is given as \cite{underlay_myopic}:
\begingroup
\begin{subequations}
\begin{align}
\max_{\vec{\pmb{\beta}}}\! & \sum_{i=1}^N\!\beta_i\log_2\!\left(\!1+\frac{(1-\beta_i)|g^i_{ss}|^2E_h^i}{\beta_i(\sigma_n^2+(|\hat{g}_{ps}^i|^2+2\varepsilon|\hat{g}_{ps}^i|+\varepsilon^2)p_p)}\right),\label{eq:myopic_obj}\\
\text{s.t.}\;\; & (1-\beta_i)(|\hat{g}_{sp}^i|^2+2\varepsilon|\hat{g}_{sp}^i|+\varepsilon^2)E_h^i\leq \beta_iP_{th},\;\forall i, \label{eq:myopic_c1}\\
& \vec{0}\preceq \vec{\pmb{\beta}}\preceq\vec{1},\label{eq:myopic_c2}
\end{align}
\end{subequations}
\endgroup
which is a convex optimization problem and can be solved using any standard convex optimization solver such as CVX \cite{cvx}.
\subsection{Offline Policy}
In the offline policy, all the channel coefficients and energy arrivals are assumed to be known apriori. The offline policy outperforms the myopic and online policies in terms of sum throughput and acts as a benchmark for these policies. The optimization problem for robust offline policy is given as \cite{my_spcom}:
\begingroup
\allowdisplaybreaks
\begin{subequations}
\begin{align}
\max_{\vec{p}_s,\pmb{\vec{\beta}}}\quad & \sum_{i=1}^NR_1(\beta_i,p_s^i), \label{eq:offline_obj}\\
\text{s.t.} \quad & \eqref{eq:c1_orig}, \, \eqref{eq:c2_orig},\, \eqref{eq:c4_orig},\, \eqref{eq:c2_new}, \label{eq:offline_c1}
\end{align}
\end{subequations}
\endgroup
which is a convex optimization problem and can be solved using any standard convex optimization solver such as CVX \cite{cvx}.
\section{Results and Discussions}
For simulation, it is assumed that the PT uses a constant power of $p_p=2$ Watt in all the slots for transmission, and $\sigma_n^2=0.1$. The number of states of the energy harvesting process is assumed to be $M_s=2$, such that $E_h$ takes values from the discrete set $\mathcal{S}=\{e_1^h=0,e_2^h=0.5\}$ depending upon the transition probability matrix
\begin{align}
\mathbf{T}=\frac{1}{2}\begin{bmatrix}
1 &1 \\ 
 1& 1
\end{bmatrix}
\label{eq:transition_prob_matrix}
\end{align}
Since we are considering discrete time SDP, the optimization variables, $\beta_i$ and $p_s^i$ are considered to be discrete with step size of 0.2.
\vspace{-1mm}
\subsection{Effect of uncertainty region radius $\varepsilon$ on secondary throughput}
\label{subsection:diff_epsilon}
\begin{figure*}[!t]
\centering
\begin{minipage}{0.3\linewidth}
\centering
  \includegraphics[width=2.3in]{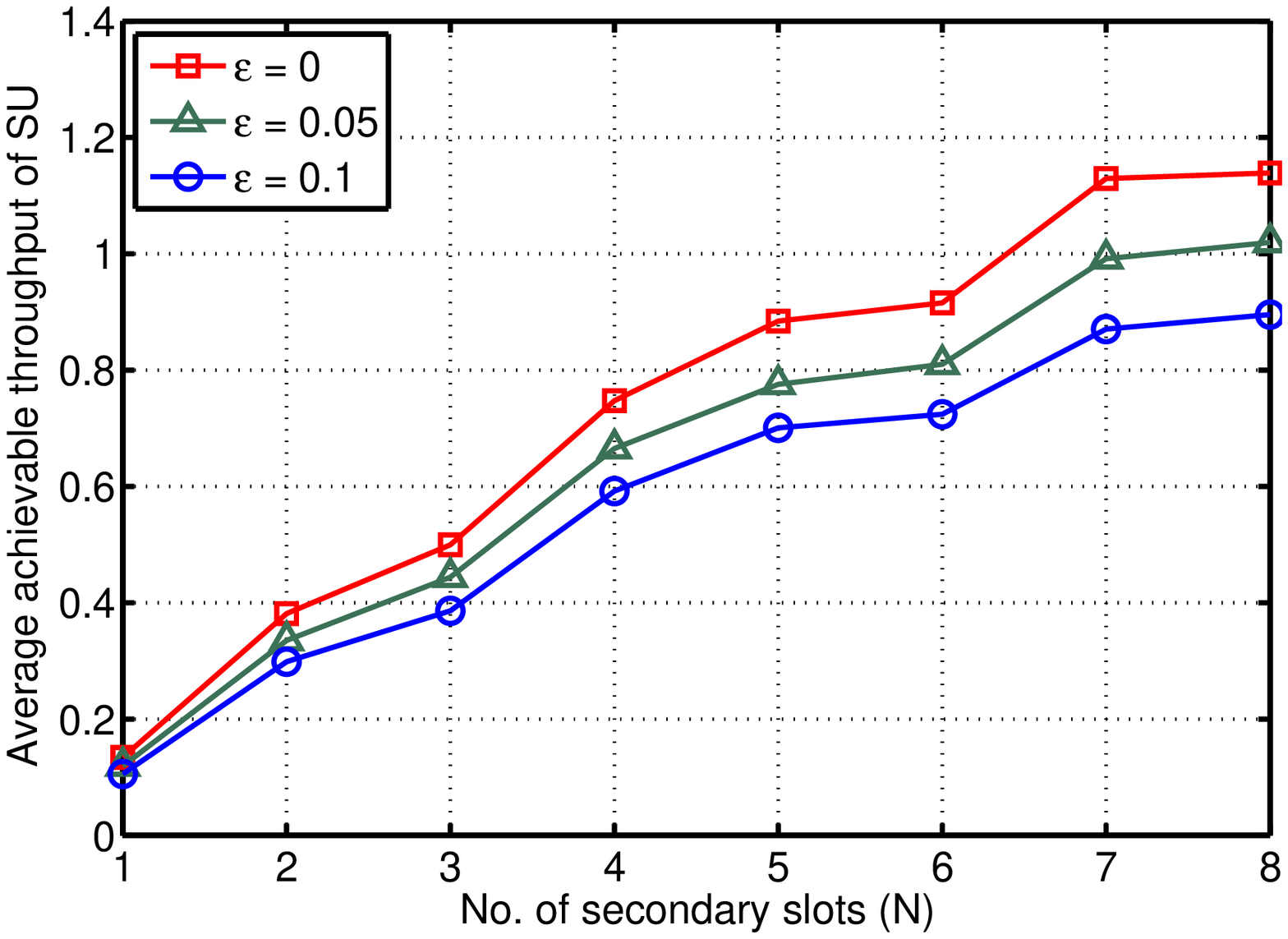}
  \caption{Average sum throughput of ST ($R_{sum}^{avg}$) versus number of slots ($N$) for different radius of uncertainty region ($\varepsilon$) under optimal online policy.}
  \label{fig:throughput_diff_channel_uncertainty_online}
\end{minipage}
\hspace{3mm}
\begin{minipage}{0.3\linewidth}
\centering
  \includegraphics[width=2.3in]{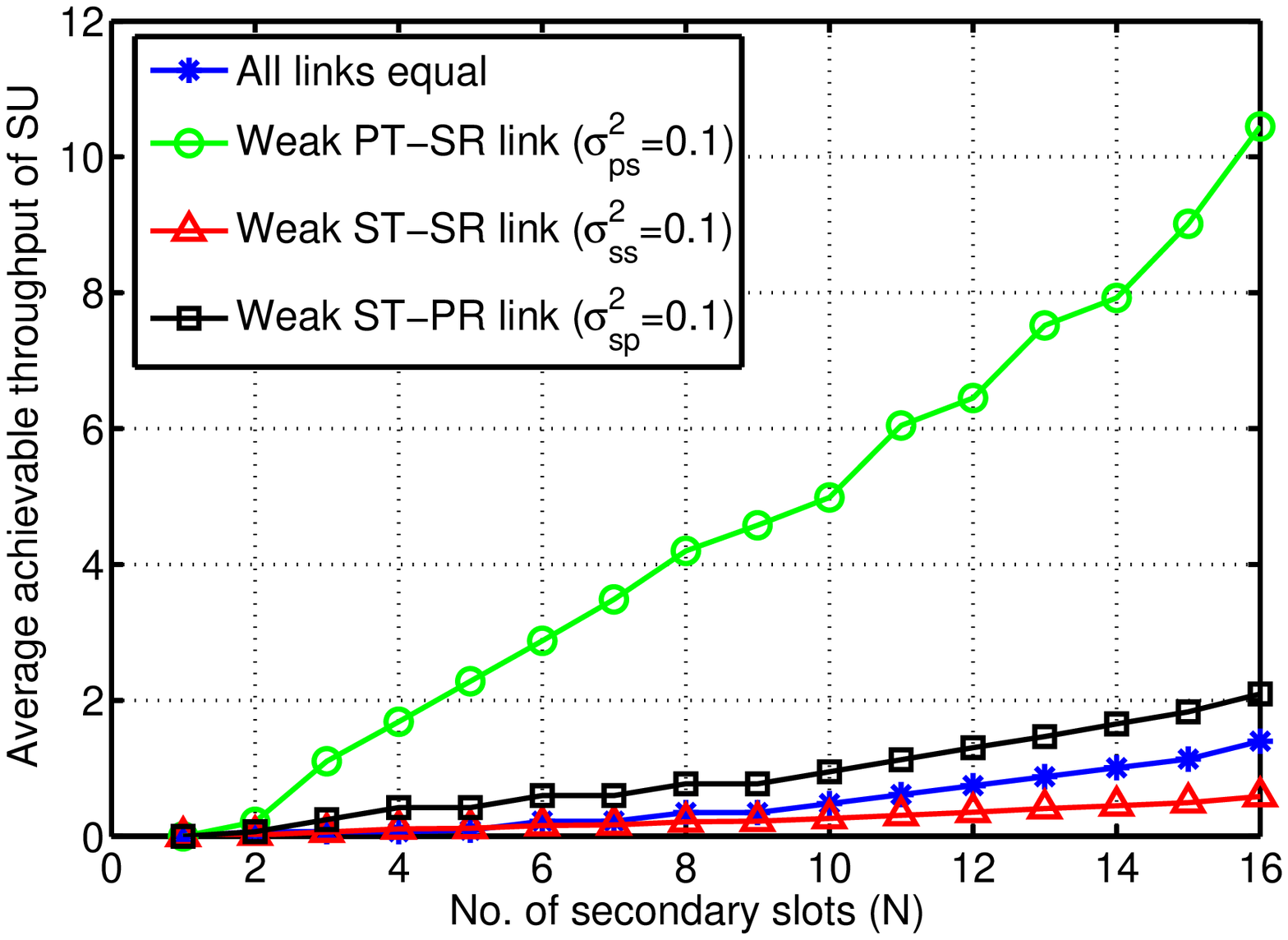}
  \caption{Average sum throughput of ST $(R_{sum}^{avg})$ versus number of slots $(N)$ for different channel conditions under optimal online policy.}
  \label{fig:throughput_diff_channel_online}
\end{minipage}
\hspace{3mm}
\begin{minipage}{0.3\linewidth}
\centering
\centering
  \includegraphics[width=2.3in]{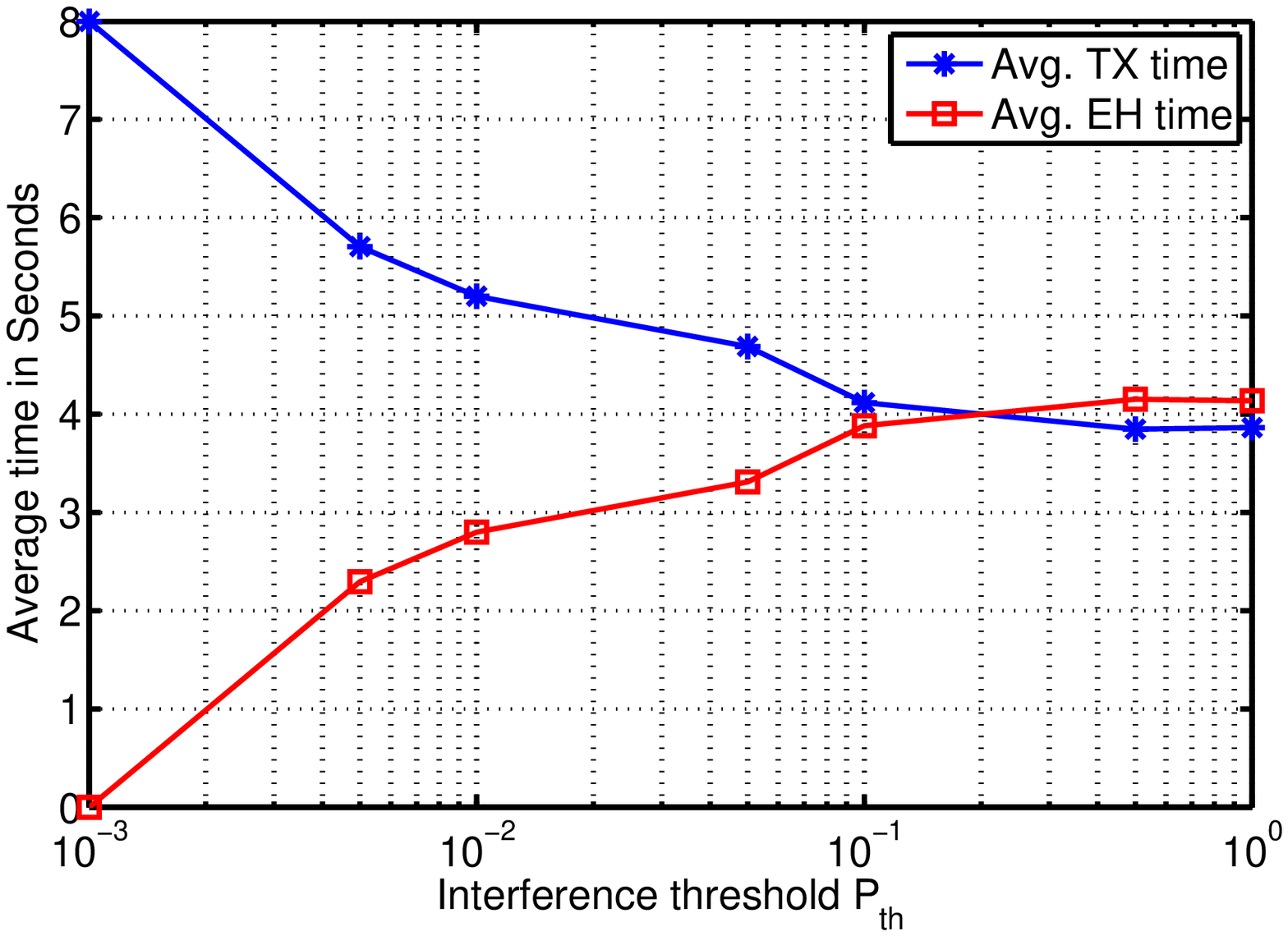}
  \caption{Average harvesting and transmission time versus interference threshold $(P_{th})$ under optimal online policy.}
  \label{fig:beta_vs_interf_online}
\end{minipage}
\vspace{-0.5cm}
\end{figure*}

Fig. \ref{fig:throughput_diff_channel_uncertainty_online} shows ST's average sum throughput ($R^{avg}_{sum}$) averaged over different channel realizations for different values of uncertainty region radius $\varepsilon$ under the optimal online time sharing policy. The variances of all the channel links are assumed to be unity, i.e., $\sigma_{pp}^2=\sigma_{ps}^2=\sigma_{sp}^2=\sigma_{ss}^2=1$, interference threshold $P_{th}=1$ Watt, and $B_{max}=1$ Joule. The effect of radius of uncertainty region on the worst case average throughput is clearly visible from the figure. As $\varepsilon$ increases, the average throughput decreases due to two reasons. First, increasing $\varepsilon$ reduces the instantaneous throughput and second, from the constraint (\ref{eq:c2_new}), increasing $\varepsilon$ puts more stringent constraint on the transmit power $\bar{\mat{p}}_s$ which in turn, reduces the average throughput.
\subsection{Effect of different channel conditions on throughput}

Fig. \ref{fig:throughput_diff_channel_online} shows ST's average sum throughput for various channel conditions under optimal online time sharing policy. For simulations, we assume that the weak and strong channel links have variance 0.1 and 1 respectively, e.g., in case of weak PT-SR link, we assume $\sigma_{ps}^2=0.1$ and $\sigma_{pp}^2=\sigma_{sp}^2=\sigma_{ss}^2=1$, $P_{th}=1$ Watt and $\varepsilon=0.05$. In case of all channel links to be equally strong, we assume $\sigma_{pp}^2=\sigma_{ps}^2=\sigma_{sp}^2=\sigma_{ss}^2=1$. From the figure, it is noticed that the secondary throughput increases as the link ST-PR  becomes weak. This results in low interference constraint at PR allowing the ST to transfer information with high power which in turn, results in higher throughput. The weak PT-SR link causes low interference at SR, therefore the throughput increases. The weak ST-SR link degrades the secondary performance because of poor channel gains. When all links are equally strong, the performance lies in between that of the weak ST-PR link and weak ST-SR link. This is because when weak ST-PR link allows ST to transmit with higher power, weak ST-SR negates this gain from weak primary interference resulting in no throughput gain.
\subsection{Effect of interference threshold $P_{th}$ on average
 harvesting/transmission time}
Fig. \ref{fig:beta_vs_interf_online} shows the variations of average harvesting/transmission time with the change in interference threshold at the SR, $P_{th}$. The plot is obtained for fixed number of secondary slots $N=8$, radius of uncertainty region $\varepsilon=0.05$, and variances of channel coefficients are assumed to be same as in section \ref{subsection:diff_epsilon}. As the value of $P_{th}$ increases, SU can transmit with more power, which can be obtained by consuming more energy in less amount of time as $P^i_s=E^i_s/\beta_i$, where $E^i_s$ is the energy consumed by ST in $i$th slot. Therefore, the harvesting time increases and transmission time decreases so that the ST can accumulate more energy and can transmit with higher power.
\subsection{Effect of battery capacity $B_{max}$}
\vspace{-3mm}
\begin{figure}[!h]
\centering
\includegraphics[width=0.75\linewidth]{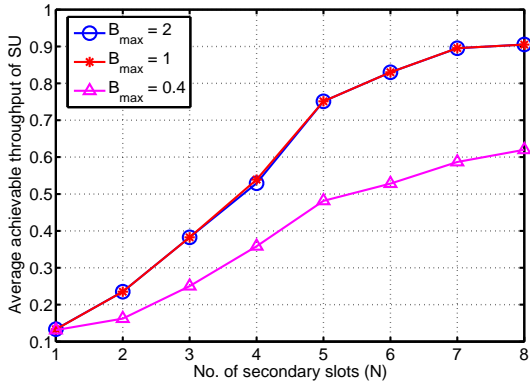}
\caption{Average sum throughput of ST $(R_{sum}^{avg})$ versus number of slots $(N)$ for different values of battery capacity $B_{max}$}
\label{fig:throughput_vs_slot_diff_B_max}
\vspace{-3mm}
\end{figure}
Fig \ref{fig:throughput_vs_slot_diff_B_max} shows the effect of battery capacity on the average achievable throughput of ST. The radius of uncertainty region is assumed to be $\varepsilon=0.05$ and all other parameters are same as in section \ref{subsection:diff_epsilon}. As we decrease the battery capacity, secondary throughput reduces. This effect of the battery capacity can be observed from constraint (\ref{eq:c1}). When battery capacity is reduced, $B_{max}$ dominates in constraint (\ref{eq:c1}) and the next state in the battery is limited by $B_{max}$, i.e., this constraint does not allow the ST to harvest the energy it needs, which reduces the throughput. As battery capacity is increased, the ST can accommodate more energy and therefore can transmit with higher power whenever channel conditions allow. Fig. \ref{fig:throughput_vs_slot_diff_B_max} shows that after a limit, further increment in battery capacity has no impact on the throughput as in this case, first term in constraint (\ref{eq:c1}) becomes dominant and $B_{max}$ has no effect on the next state of the battery.
\subsection{Nature of harvested and consumed energies}
\vspace{-3mm}
\begin{figure}[!h]
\centering
\includegraphics[width=0.75\linewidth]{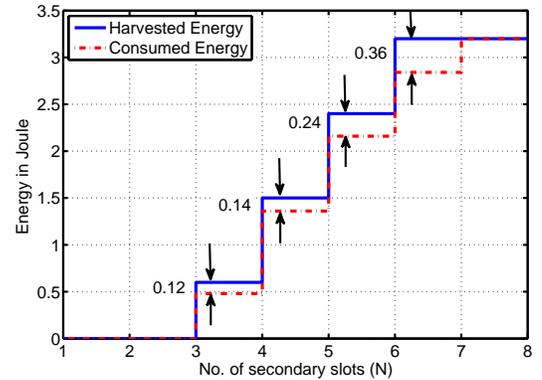}
\caption{Energy versus number of slots $(N=7)$ under optimal online policy}
\label{fig:consumed_and_harvested_energy_online}
\vspace{-3mm}
\end{figure}
Fig. \ref{fig:consumed_and_harvested_energy_online} shows the nature of harvested and consumed energy with number of secondary slots $N$. All the simulation parameters are kept same as in section \ref{subsection:diff_epsilon} and $\varepsilon=0.05$. From the figure, it is clear that in order to satisfy the energy causality constraint in (\ref{eq:c1_orig}), consumed energy always remains less than or equal to harvested energy and, the remaining energy is less than the maximum storage capacity $B_{max}=1$ Joule. Since it is a joint optimization of time and energy and it will not harvest energy which it can't use. Therefore at the end of the transmission, all the harvested energy is consumed under the optimal online policy.
\subsection{Performance comparison between the optimal online, offline and myopic time sharing policies}
\begin{figure}[!h]
\centering
\includegraphics[width=0.75\linewidth]{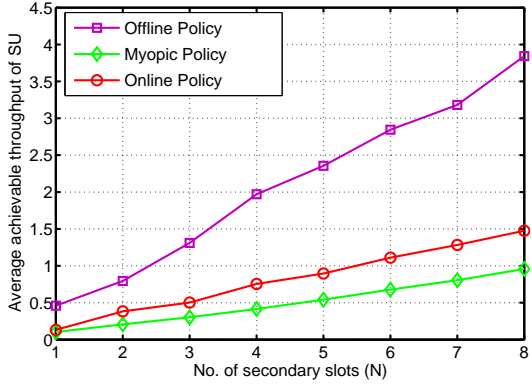}
\caption{Average sum throughput of ST $(R_{sum}^{avg})$ versus number of slots $(N)$  for optimal online and offline policies}
\label{fig:online_vs_offline_and_myopic_full_CSI}
\vspace{-5mm}
\end{figure}
Fig. \ref{fig:online_vs_offline_and_myopic_full_CSI} shows the comparison of online, offline and myopic policies in terms of average sum throughput of ST. All simulation parameters are same as in section \ref{subsection:diff_epsilon}. The offline and myopic policies considered for comparison are adopted from \cite{underlay_myopic} and \cite{my_spcom}, and modified slightly in accordance with our system model. The figure shows that the average sum throughput of ST for online policy lies in between the offline and myopic policies. Since in the offline policy, all the channel gains are assumed to be known apriori, the ST obtains the optimal policy before the transmission starts and therefore, achieves much higher throughput. Therefore, it acts as a benchmark for the online transmission policy. On the other hand, the myopic policy tries to maximize immediate throughput and consumes all the harvested energy in the same slot and therefore performs worse than the online policy.

\section{Conclusions}
%
We proposed a robust online time sharing policy that maximizes the throughput of SU under energy arrival and channel uncertainties. The proposed policy jointly optimizes the time sharing between the \textit{harvesting phase} and \textit{transmission phase}, and the transmit power of ST. The results show that our proposed policy outperforms the myopic policy and unlike offline policy, it does not require any prior information of energy arrivals and channel gains. However, the computational complexity of our policy is more than that of the myopic and offline policies as the SDP suffers from \textit{curse of dimensionality}. 

\appendix{From the optimization problem (\ref{eq:obj_orig})-(\ref{eq:c4_orig}), the problem (\ref{eq:obj_new})-(\ref{eq:c2_new}) is obtained by solving:
\begin{align}
\min_{|\Delta g_{ps}^i|\leq \varepsilon} &\sum_{i=1}^{N} \beta _{i}\log_{2}\left ( 1+\frac{\left | g_{ss}^{i} \right |^{2}p_s^{i}}{\sigma_n^2+\left(\left| \hat{g}_{ps}^{i}+\Delta g^i_{ps} \right|^{2}\right)p_{p} } \right )\label{eq:appendix_obj}
\end{align}

Using the triangle inequality, we have
\begin{align*}
\left|\hat{g}^i_{ps}+\Delta g^i_{ps}\right| \leq \left|\hat{g}^i_{ps}\right|+\left|\Delta g^i_{ps}\right|
\end{align*}
squaring both sides gives
\begin{align}
\left|\hat{g}^i_{ps}+\Delta g^i_{ps}\right|^2& \leq \left|\hat{g}^i_{ps}\right|^2+\left|\Delta g^i_{ps}\right|^2+2\left|\hat{g}^i_{ps}\right|\left|\Delta g^i_{ps}\right|\nonumber\\
&\leq \left|\hat{g}^i_{ps}\right|^2+\varepsilon^2+2\varepsilon\left|\hat{g}^i_{ps}\right| \label{eq:ineq_ps}
\end{align}
using the inequality (\ref{eq:ineq_ps}), the closed form solution of (\ref{eq:appendix_obj}) is given as
\begin{align}
\sum_{i=1}^N\beta_i\log_2\left(1+\frac{\left|g^i_{ss}\right|p_s^i}{\sigma_n^2+\left(\left|\hat{g}^i_{ps}\right|^2+2\varepsilon\left|\hat{g}^i_{ps}\right|+\varepsilon^2\right)p_p}\right)
\end{align}
similarly, we have
\begin{align}
\left|\hat{g}^i_{sp}+\Delta g^i_{sp}\right|^2& \leq \left|\hat{g}^i_{sp}\right|^2+\varepsilon^2+2\varepsilon\left|\hat{g}^i_{sp}\right| \label{eq:ineq_sp}
\end{align}
Therefore, we replace the constraint (\ref{eq:c3_orig}) with
\begin{align}
(| \hat{g}_{sp}^{i}|^2 +2\varepsilon |\hat{g}_{sp}^{i}|+\varepsilon ^2)p_s^{i}\leq P_{th},\;\forall i
\end{align}
which means that the worst case interference should also be less than or equal to the interference threshold at PR, $P_{th}$.
}

\bibliographystyle{ieeetr} 
\bibliography{references}

\end{document}